\documentclass[mathleft
]{an}
\usepackage{graphicx}
\usepackage{times}
\begin{document}
\Yearpublication{2010}%
\Yearsubmission{2010}%
\Month{00}%
\Volume{000}%
\Issue{000}%

\title{Asteroseismic modelling of Procyon A: Preliminary results}
\author{G. Do\u{g}an\inst{1}\fnmsep\thanks{\email{gulnur@phys.au.dk}\newline}
A. Bonanno \inst{2} T. R. Bedding \inst{3} T. L. Campante \inst{1,4}
J. Christensen-Dalsgaard \inst{1} \\ \and H. Kjeldsen \inst{1}}
\titlerunning{Modelling Procyon A}
\authorrunning{G. Do\u{g}an et al.}
\institute{ Department of Physics and Astronomy, Aarhus University,
Ny Munkegade, DK-8000, Aarhus C, Denmark \and Catania Astrophysical
Observatory, Via S.Sofia 78, 95123, Catania, Italy \and School of
Physics A29, University of Sydney, NSW 2006, Sydney, Australia \and
Centro de Astrof\'{\i}sica da Universidade do Porto, Rua das
Estrelas, 4150-762 Porto, Portugal}

\received{XX} \accepted{XX}


\abstract{%
  We present our preliminary results of the modelling of the F5
star Procyon A. The frequencies predicted by our models are compared
with the frequencies extracted through a global fit to the power
spectrum obtained by the latest ground-based observations, which
provides two different mode identification scenarios.}

\maketitle

\vspace{-12pt}
\section{Introduction}
\vspace{-6pt}

\sloppy

Procyon A is a member of a binary system with a white dwarf
companion, Procyon B. It is one of the very bright stars to the
naked eye, and hence it has been very attractive for the observers.
The observational constraints which we adopted are summarized in
Section 2. It has also been of asteroseismic interest for a long
while (see Arentoft et al. 2008, for a summary), with a solar-like
power excess in its spectrum first reported by Brown et al.~(1991).
However, there has been no agreement on the individual oscillation
frequencies. Several authors have investigated the structure and
evolution of Procyon A through an asteroseismic approach (e.g.,
Guenther \& Demarque 1993; Barban et al.~1999; Chaboyer et al.~1999;
Di Mauro \& Christensen-Dalsgaard 2001; Eggenberger et al.~2005;
Provost et al.~2006, Bonanno et al.~2007), but there has been a need
for more accurate frequencies. Recently, the star was observed
through a multi-site campaign by eleven telescopes for more than
three weeks (Arentoft et al.~2008). The frequency analysis is
described by Bedding et al.~(2010). They presented results from
different approaches of frequency extraction: Iterative sine-wave
fitting and global fitting to the power spectrum. In the former
method, a sine wave is fitted to each mode one after the other while
the corresponding sinusoid is subtracted from the time series at
each step. This is repeated until the signal to noise ratio of the
remaining power is lower than a given threshold. This method was
used for frequency extraction of ground-based radial velocity data
before (see, e.g., the analysis on solar-like star $\beta$ Hyi by
Bedding et al.~(2007)). In the latter method, the goal is to find an
overall fit to the power spectrum for all the frequencies, mode
heights, and linewidths simultaneously, using some prior knowledge
of oscillation properties as constraints. A similar implementation
of this method was previously applied to space-based data (see,
e.g., frequency analysis of CoRoT star HD$\:$49933 by Benomar et
al.~(2009)). In this work, we adopted this Bayesian approach which
provided us with two mode identification scenarios, referred to as
Scenario A and B, with different posterior odds (for a detailed
discussion, see Bedding et al.~2010). We chose the output of this
analysis in order to test both scenarios.  Note that Scenario B was
favoured by Bedding et al. (2010; see that paper for a discussion),
but here we test both scenarios.

\fussy

\vspace{-6pt}
\section{Methods and Tools}
\vspace{-6pt}

\sloppy

We have adopted the following properties for the position of the
star in the H-R Diagram:
$T_{\rm{eff}}\:$=$\:$6530$\:\pm\:90\:\rm{K}$ (Fuhrmann et al.~1997)
and log($L/L_{\odot}$)$\:$=$\:$0.85$\:\pm\:$0.06 (Steffen 1985). We
note that there are several revised values for luminosity in the
literature, such as log$(L/L_{\odot})\:$=$\:$0.840$\:\pm\:$0.018
derived by Jerzykiewicz \& Molenda-\.{Z}akowicz (2000) using the
Hipparcos parallax and total absolute flux; however, we chose to
scan a wider range, which largely covers most of the revised values.
A similar argument applies also to the choice of the effective
temperature. We have not put an additional constraint on the radius
for the time being, although we do compare the stellar mean density
inferred from the analysis with the value
0.172$\:\pm\:$0.005$\:\rho_{\odot}$, obtained from the measured
radius using the angular diameter 5.404$\:\pm\:0.031\:\rm{mas}$
(Aufdenberg et al.~2005) and the revised Hipparcos parallax
($284.56\:\pm\:1.26\:\rm{mas}$\footnote{Note that there is an error
in Bedding et al. 2010 (Section 9); the value they give for the
revised parallax is actually the original one. We also note that the
uncertainty on the revised parallax is larger than the original but
is presumably more reliable.}, van Leeuwen 2007), together with the
adopted mass $1.463\:\pm\:0.033\:M_{\odot}$, which is the mean of
the two different astrometric determinations (Girard et al.~2000,
Gatewood and Han 2006). For the metallicity of the star, we allowed
a wide range covering the $0.05\:\rm{dex}$ iron deficiency suggested
by Allende Prieto et al.~(2002), and we used the solar mixture from
Grevesse \& Noels (1993).

We have computed stellar models with two different evolutionary
codes: ASTEC (Aarhus STellar Evolution Code) (Christensen-Dalsgaard
2008a) and GARSTEC (Garching Stellar Evolution Code) (Weiss \&
Schlattl 2008). The method we used is to compute several grids of
standard models scanning through a parameter space formed by varying
the mass, the initial metallicity at the stellar surface,
$Z_{\rm{i}}/X_{\rm{i}}$, and the mixing-length parameter, $\alpha$,
where the mixing length is defined as $\ell\:$=$\:\alpha H_{p}$,
$H_{p}$ being the pressure scale height. So far, we have searched
within the following ranges:
$M$$\:$=$\:$1.42$\:$--$\:$1.52$\:M_{\odot}$,
$Z_{\rm{i}}/X_{\rm{i}}$$\:$=$\:$0.0204$\:$--$\:$0.0245,
$Y_{\rm{i}}$$\:$=$\:$0.26$\:$--$\:$0.31, and
$\alpha$$\:$=$\:$1.6$\:$--$\:$1.9. Here $X_{\rm{i}}$, $Y_{\rm{i}}$,
and $Z_{\rm{i}}$, are the initial mass fractions of hydrogen,
helium, and the elements heavier than helium, respectively. We have
computed the models without taking into account diffusion,
convective overshooting, or rotation.

The Aarhus adiabatic pulsation package (ADIPLS)
(Christensen-Dalsgaard 2008b) has been used to calculate the
frequencies of the models having properties that are in agreement
with the observations. We have then compared the model frequencies
with the observed frequencies. We selected the models that minimize
the following $\chi^2$:
$$
\chi^{2}=\sum_{n,l}\left(\frac{\nu_{l}^{\rm{obs}}(n)-\nu_{l}^{\rm{model}}(n)}{\sigma(\nu_{l}^{\rm{obs}}(n))}\right)^{2},
$$ where $\nu_{l}^{\rm{obs}}(n)$, and $\nu_{l}^{\rm{model}}(n)$ are the
model, and the observed, frequencies with spherical degree $l$ and
radial order $n$, and $\sigma(\nu_{l}^{\rm{obs}}(n))$ represents the
uncertainties in the observed frequencies.


\fussy

\section{Results}

\sloppy

The results from the two different stellar evolution codes are
similar; hence we present some of the selected models computed with
ASTEC. The so-called \'{e}chelle diagrams of the best models for
both of the scenarios, chosen without applying any near-surface
corrections are shown in Figs 1 and 2, with their parameters
summarized in Table 1. One plots the \'{e}chelle diagrams using the
frequency modulo the large frequency separation, $\Delta\nu$, in the
horizontal axis. In order to allow easy comparison with the diagrams
shown by Bedding et al.~(2010), we use the same value,
$\Delta\nu\:$=$\:56\,\mu$Hz, in our diagrams.

\fussy

\begin{figure}[htbp]
\begin{center}
\includegraphics[scale=0.3,angle=90]{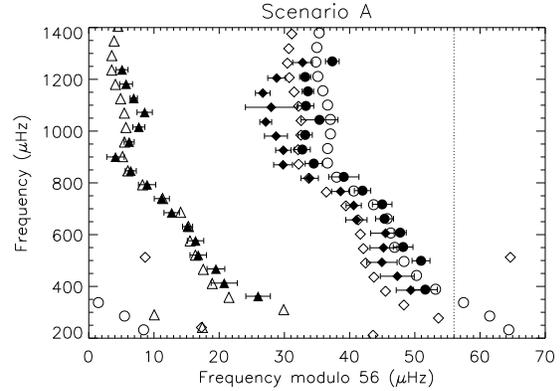}
\end{center}
\caption{\'{E}chelle diagram of the selected model for Scenario A.
Open symbols represent the model frequencies, while the filled
symbols with the uncertainties correspond to the frequencies
extracted from the observations. Circles, triangles, and diamonds
are used for the modes with spherical degree $l$= 0, 1, and 2,
respectively. The vertical dot-dashed line corresponds to the value
of $\Delta\nu$.} \label{Figure 1}
\end{figure}

\begin{figure}[htbp]
\begin{center}
\includegraphics[scale=0.3,angle=90]{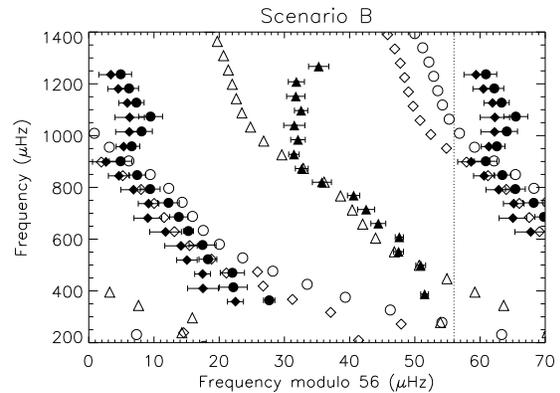}
\end{center}
\caption{\'{E}chelle diagram of the selected model for Scenario B.
Symbols are used in the same way as in Fig. 1.} \label{Figure 2}
\end{figure}

\begin{table}[h]
\begin{center}
 \caption{Parameters of the best models}
\begin{tabular}{@{}c|c|c}
\hline
Parameter & Scenario A  & Scenario B \\
\hline
$M/M_{\odot}$ & 1.50 & 1.50 \\
$Z_{\rm{i}}/X_{\rm{i}}$& 0.0235 & 0.0245  \\
$Y_{\rm{i}}$ & 0.266 & 0.290 \\
Age (Gyr) & 1.83 & 1.51 \\
$R/R_{\odot}$ & 2.058 & 2.067 \\
$\rho/\rho_{\odot}$ &0.1721& 0.1698\\
$L/L_{\odot}$ & 6.565 & 7.286 \\
$T_{\rm{eff}}$(K) & 6446 & 6603\\
$\alpha$&1.8&1.6\\
$X_{c}^{*}$ & 0.1585 & 0.1995 \\
\hline
$\chi^{\scriptscriptstyle 2}$ & 3.79 & 24.46 \\
\hline
\end{tabular}
\end{center}
$^{*}$ \footnotesize Mass fraction of hydrogen remaining in the
centre of the star
\end{table}
The behaviour of the frequency differences between the models and
the observations (shown in Figs 3 and 4) are quite different from
that in the Sun. Kjeldsen et al.~(2008) showed that the difference
between observed and model frequencies of the Sun can be fitted by a
power law, which can also be employed to correct the model
frequencies for near-surface effects in other solar-like stars, such
as $\beta$ Hyi and $\alpha$ Cen A. However, in Procyon, we cannot
justify the application of such a surface correction to yield a
significant improvement in the fit, since the frequency differences
do not follow the power-law behaviour (see also Figs 16 and 17 of
Bedding et al.~2010).

\begin{figure}[h]
\begin{center}
\includegraphics[scale=0.25]{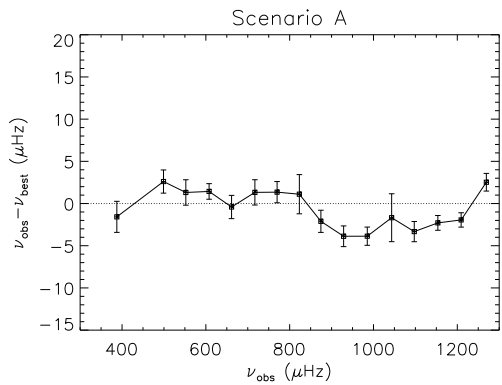}
\end{center}
\caption{The difference between the radial ($l=0$) frequencies from
the observations and the selected model for Scenario A. The
indicated uncertainties are those from the data analysis.}
\label{Figure 3}
\end{figure}

\begin{figure}[h]
\begin{center}
\includegraphics[scale=0.25]{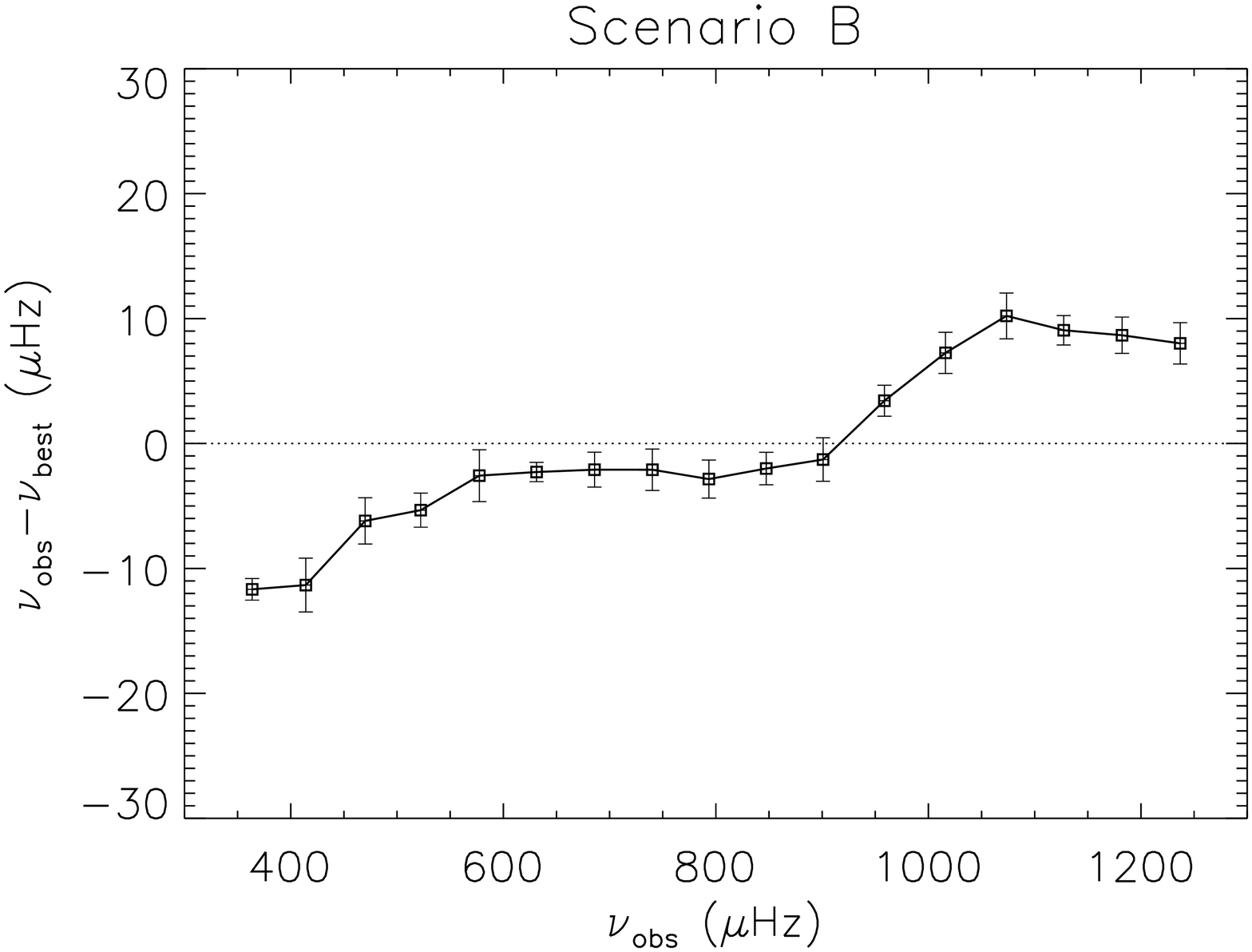}
\end{center}
\caption{Same as Fig. 3, but for Scenario B} \label{Figure 4}
\end{figure}


\section{Discussion and Conclusion}

We can argue, if Scenario A is the correct one, that the predictions
of the stellar evolutionary models match the observations quite
well; however, a surface correction for the model frequencies seems
not to be needed, unlike in the solar case. Therefore, the effects
of different near-surface properties on the high frequencies might
be cancelling out in Procyon; this deserves further investigation.

\sloppy

If, on the other hand, Scenario B is correct, there seems to be no
good agreement between the models and the observations, even in the
low-frequency region, which means that there is something
incompatible in the structure of the models.

We have presented preliminary results from our on-going work. To
come to a conclusion we need to extend our analysis. Effects of
inclusion of overshooting and use of different treatments of
convection should be analysed. In addition, the suspected mixed mode
reported by Bedding et al.~(2010) could help us distinguish between
the two scenarios, and put further constraints on the age, and the
chemical composition.

Although from the point of view of modelling, Scenario A seems to be
less problematic, we cannot yet strongly favour either of the
scenarios; therefore, it is difficult to set accurate constraints on
the stellar properties. Nevertheless, either of the cases suggests
that Procyon is quite different from the Sun, which provides a very
good opportunity to test our understanding of stellar structure and
evolution.

\acknowledgements  This work was supported by the European Helio-
and Asteroseismology Network (HELAS), a major international
collaboration funded by the European Commission's Sixth Framework
Programme. GD, JC-D, and HK acknowledge financial support from the
Danish Natural Science Research Council. GD would like to thank
Pierre-Olivier Quirion for providing a parallel-programming code for
faster evolution computation.

\fussy


\end{document}